\documentclass[a4paper,12pt]{article}
\usepackage{graphicx}
\begin{document}
\begin{center}
\Large\bf
Diffusion and local deconfinement in relativistic systems\\[2.1cm]
\large
Georg Wolschin\footnote{E-mail: wolschin@uni-hd.de\hspace{5cm}
http://wolschin.uni-hd.de}\\[.8cm]
\normalsize\sc\rm
Institut f\"ur Theoretische Physik der Universit\"at,
D-69120 Heidelberg, Germany\\[2cm]
\end{center}
\bf
\rm
\bf{Abstract:}
\normalsize\sc\rm
In relativistic systems at RHIC energies, new deconfinement signatures
emerge and indicate an increasingly clear separation between soft
hadronic processes, and hard partonic interactions
in a locally deconfined subsystem. Here the emphasis
is on longitudinal variables, namely, net-baryon rapidity distributions.
As described in a Relativistic Diffusion Model,
they change from bell-shaped at the lower to double-humped
at the higher SPS-energy, but do not  reach local statistical equilibrium.
At $\sqrt{s_{NN}}$ = 200 GeV in the Au-Au system, however, they are shown to
consist of three components. In addition to the nonequilibrium
contributions, a third fraction close to midrapidity containing
$Z_{eq}\simeq 22$
protons reaches local statistical equilibrium in a
discontinuous transition. It may be associated with a deconfinement of the
participant partons and thus, serve as a signature
for Quark-Gluon Matter formation.\\[.4cm]
\bf{PACS numbers:}
\normalsize\sc\rm
25.75.-q, 24.60.Ky, 24.10.Jv, 05.40.-a
\newpage
\section{Introduction}
Considerable progress has recently been made in the long-standing
attempts \cite{nan03} to recreate a quark-gluon plasma under laboratory
conditions by means of relativistic heavy-ion collisions.
Quark Matter is believed to constitute an important intermediate
state of the very early
universe. There it is in thermal equilibrium, but expands, cools and ends in the
most dramatic event of a quark-hadron phase
transition - a thermal confinement transition - at about 10$\mu$s:
Quarks and gluons condense to form a
gas of nucleons and light mesons, the latter decaying
afterwards.

As the temperature drops to about 170 MeV, the hadron gas becomes
sufficiently dilute, and the hadron abundances for
different particle species remain fixed ("chemical freezeout").
The phase transition has thus set the stage for the subsequent
primordial synthesis of light nuclei (d, He-3, He-4, Li-7)
at times t$\simeq$1s, and temperatures T$\simeq$1 MeV in the evolution
of the early universe \cite{djs03}.

The attempt to investigate the quark-hadron phase transition in reverse
order, as a deconfinement transition, together with the subsequent
confinement transition in the laboratory by means of
relativistic heavy-ion collisions experiences severe difficulties.
The system has to be sufficiently extended, small collision partners
such as protons or deuterons
are not suitable. The energy density has to reach values above
the critical one (about 1.5 GeV/fm$^{3}$) for the transition to
occur.

The typical time scale for relativistic heavy-ion collisions in the laboratory
is only about 10$^{-23}$s - to
be compared with the much larger time scale of the cosmological QCD
transition of
 10$^{-5}$s. Hence, it can not be expected that thermal
equilibrium, which governs the physical description of the early
universe, remains a valid concept for theoretical models of
relativistic heavy-ion collisions. Whereas particle
abundances have been shown to be described rather
accurately by phase-space ("thermal") models \cite{hag65,bec01,bra01},
this does not necessarily imply that the
system has reached, or gone through, thermal equilibrium. Instead, one has to
look for stages of \it{local} \sc\rm kinetic equilibrium in the short time
evolution
of the system, and for the possibility that the deconfinement
transition occurs in such a stage of local thermal equilibrium,
affecting only a relatively small number of nucleons in a
relatively big system.\\

In the fixed-target experiments at the SPS with heavy systems -
in particular, with the Pb-Pb system at
$\sqrt{s_{NN}}$ = 17.3 GeV - a number of possible phase-transition
signatures such as strangeness enhancement, and excess of dileptons with
invariant mass below that of the $\rho$-meson had been discussed.
The most promising signal, namely, the suppressed production of the
${J/\Psi}$-meson in the presence of a quark-gluon plasma due to
vanishing string tension and screening, had been predicted by theorists
\cite{mat86} and
identified at the SPS in heavy systems, but since it could also be caused
by hadronic final state interactions (nuclear absorption) it seemed
not fully convincing.

Whether the "extra suppression" which was then detected
in the Pb-Pb system at $\sqrt{s_{NN}}$ = 17.3 GeV and which could
not easily be accounted for
by absorption constitutes a qgp-signature
ist still a matter of debate. At the RHIC energy of 200 GeV per
particle, the PHENIX collaboration has presented preliminary results
for the ${J/\Psi}$-meson \cite{rey03} showing a slight suppression.
In view of the large error bars, however, this is not yet conclusive
either, one has to wait for more precise data.

In a probably more promising effort, the four RHIC Au-Au experiments
have carefully investigated the particle production
in central collisions at high transverse momenta. When compared to
p-p data that are scaled with the number of binary collisions, a
significant suppression of the produced hadrons is found, which is
interpreted as a final state effect of the produced dense medium - and
possibly, of a quark-gluon plasma. The effect may be due to
"jet quenching": Energetic partons
traversing the dense medium lose energy or are completely
absorbed, and the remaining observed hadronic jets are
mostly created from partons produced near the surface and directed
outwards.

The effect is not observed
(instead, the inclusive yield is slightly enhanced) in the lighter d-Au system,
where compression and heating is much less
pronounced, and where qgp-formation is therefore unlikely
\cite{bac03,adl03,ada03,ars03}. Moreover, back-to-back pairs are also
strongly suppressed in central Au-Au for similar reasons, whereas
near-side pairs exhibit jet-like correlations that are similar to
p-p- and d-Au-results, also pointing towards jet absorption in the dense
(qgp?) medium \cite{cadl03}.

Whereas these results focus essentially on transverse variables,
longitudinal variables also offer very interesting
conclusions regarding hadronic vs. partonic interactions, and the
possibility of qgp formation. In this contribution,
I shall therefore concentrate on longitudinal variables - in particular,
rapidity distributions of net baryons. The analysis is
based on a Relativistic Diffusion Model (RDM)
that allows to deal with analytical solutions, rather than numerical
codes that often provide little insight into the physical assumptions.
\section{Relativistic diffusion model}
The Relativistic Diffusion Model emphasizes the
nonequilibrium-statistical
features of relativistic heavy-ion collisions. It also encompasses
kinetic (thermal) equilibrium of the system for times that are
sufficiently larger than the relaxation times of the relevant
variables such as transverse energy E$_{\perp}$ or rapidity y.

A first (linear) version of the RDM had been proposed in 1996 and
applied successfully to the analysis of AGS- and SPS-data, with an emphasis
on transverse energy distributions integrated over all particle species,
and the mean value of the rapidity as function of transverse
energy \cite{wol96}. Although transverse energy spectra of produced particles
turn out to be close to thermal equilibrium, some deviations from
equilibrium appear in the transverse variables. Based on analytical
solutions of a transport equation, accurate
predictions of transverse energy spectra were made, such as in case
of the Pb-Pb system at the highest SPS energy.

Distributions of longitudinal variables are of greater interest in the
RDM-approach since they remain farther away from thermal equilibrium.
This is particularly true for the rapidity
y=1/2 ln((E+p)/(E-p)), which
is the Lorentz-invariant counterpart of the velocity in the beam
direction at relativistic energies.
Hence, I have focussed the RDM in 1999 on rapidity distributions
\cite{wol99}. For net (participant) baryons, $\delta$-function initial
conditions corresponding to the beam rapidities are appropriate, and
analytical solutions of the rapidity transport equation can be
compared with data for net-proton rapidity spectra.

Although the initial conditions are less staightforward for produced
hadrons, Biyajima et al. have started in 2002 \cite{biy02} to use the
analytical RDM - which
they had developed independently from, but with exactly the same
result as in \cite{wol99,wols99} - for produced hadrons. Comparing to a vast
amount of RHIC data for produced charged hadrons (PHOBOS and BRAHMS)
at both 130 GeV and 200 GeV
center-of-mass energy per particle pair, they obtain high-precision
fits of the data with adjusted values of the friction coefficient
(rapidity relaxation time in my terminology) and the variance (rapidity
diffusion coefficient) \cite{biy02,biy03}. With the RDM-approach, they are also
able to relate pseudorapidity and rapidity distributions to each
other, and to establish scaling of the charged-particle rapidity
distributions with the number of charged hadrons (rather than the
number of participants, or binary collisions).

Based on rapidity diffusion coefficients that are not fitted, but
instead calculated analytically
using a dissipation-fluctuation theorem in
the weak-coupling limit (where the time between two subsequent interactions
is large compared to the duration of an individual interaction), I
have in 1999 obtained good
RDM-results for net proton distributions at the low SIS
energies of about 1 GeV per particle. However, this limit is not
attained at AGS and SPS energies, where progressively larger
deviations between RDM weak-coupling result and data occur \cite{wols99}.
These deviations have been confirmed in an independent numerical
calculation by Lavagno in 2002 \cite{lav02}.

Two solutions of this problem have been offered so far. In a
strong-coupling treatment, the diffusion coefficient D$_{y}$ in rapidity space
becomes time-dependent \cite{wols99,wol02}. For certain
parametrizations of D$_{y}$(t), analytical solutions of the RDM are
still possible. As a simple substitute in comparisons with data,
the enhancement factor due to multiparticle (collective) effects may
be determined from the deviation between weak-coupling solution and
data. Typical results for net-proton rapidity distributions at the
lower SPS beam momentum of 40 GeV/c, the higher SPS beam momentum of
158 GeV/c per particle, and RHIC (100 GeV/c per particle in each beam)
are shown in Fig.1.

Alternatively, one may resort to non-extensive statistics, with the
underlying relativistic diffusion equation in rapidity space becoming
nonlinear. This approach has been used by Alberico et al. in 2000
for transverse mass spectra and transverse momentum fluctuations at
SPS energies, assuming that they were in statistical
equilibrium \cite{alb00}. In view of the discrepancy between the
nonequilibrium weak-coupling result and the SPS Pb-Pb data,
the approach has
been extended to the nonequilibrium situation in rapidity space, where
Tsallis' nonextensivity parameter \cite{tsa88} q has been determined
from the 158A GeV Pb-Pb data \cite{lav02}. Whereas this may well be
a reasonable phenomenological parametrization of the data, it appears to
cover up the multiparticle effects which emerge explicitely in the
linear approach when comparing data and weak-coupling
dissipation-fluctuation theorem.

Subsequently both the linear (extensive) and the nonlinear
(non-extensive) approach to nonequilibrium processes in
relativistic many-body systems are pursued further, with an
emphasis on the most recent high-energy results which are investigated
experimentally at RHIC. I concentrate here on net baryons, and refer the
reader to Biyajima et al. for produced hadrons in the linear RDM
\cite{biy02,biy03,biya03}.
\section{Net baryon rapidity spectra}
Rapidity distributions of participant (net) baryons are very sensitive
to the dynamical and statistical properties of nucleus-nucleus
collisions at high energies. Recent results for net-proton rapidity spectra in
central Au+Au collisions at the highest RHIC energy
of $\sqrt{s_{NN}}$ = 200 GeV show
an unexpectedly large rapidity density
at midrapidity. The BRAHMS collaboration finds \cite{lee02}
dN/dy = 7.1 $\pm$ 0.7 (stat.) $\pm$ 1.1 (sys.) at y = 0.

The  $\Lambda,\bar \Lambda$ feed-down corrections reduce this
yield by 17.5 per cent \cite{lee02} when performed in
accordance with the PHENIX $\Lambda-$results \cite{adc02} at 130 GeV,
but the amount of stopping remains significant, although
a factor of about 4 smaller as compared to
Pb-Pb at the highest SPS energy. (A corresponding STAR result \cite{adl01}
for y=0 at 130 GeV does not yet include the feed-down correction). Many of
the available numerical microscopic models encounter
difficulties to predict the net-proton yield in the
central midrapidity valley of the distribution,
together with the broad peaks at the
detected positions.

Here I interpret the data in the
nonequilibrium-statistical Relativistic Diffusion Model.
The net baryon rapidity distribution at RHIC energies
emerges from a superposition of the beam-like nonequilibrium
components that are broadened in rapidity space through
diffusion due to soft (hadronic, low $p_{\perp}$) collisions and
particle creations, and a statistical
equilibrium (thermal) component at midrapidity
that arises from hard (partonic, high $p_{\perp}$) processes
\cite{wol03}.

At RHIC energies, the underlying distribution
functions turn out to be fairly well separated in rapidity space.
Since the transverse degrees of freedom are in (or very close to)
thermal equilibrium, they are expected to decouple from the
longitudinal ones. The time evolution of the distribution
functions is then governed by a Fokker-Planck \cite{kam81}
equation (FPE) in rapidity space
\cite{wol99,wols99,lav02,alb02,ryb02,biy02,wol03}. In the more general case of
nonextensive (non-additive) statistics \cite{tsa88} that accounts
for long-range interactions and violations
of Boltzmann's Sto{\ss}zahlansatz \cite{lav02,alb02,alb00} as well as for
non-Markovian memory (strong coupling) effects \cite{wol02,alb00},
the resulting FPE for the rapidity distribution function
R(y,t) in the center-of-mass frame is
\cite{wol03}

\begin{equation}
\frac{\partial}{\partial t}[ R(y,t)]^{\mu}=-\frac{\partial}
{\partial y}\Bigl[J(y)[R(y,t)]^{\mu}\Bigr]+D(t)
\frac{\partial^2}{\partial y^2}[R(y,t)]^{\nu} .
\label{fpe}
\end{equation}\\
Here, the rapidity diffusion coefficient D(t) may in general be
time-dependent, although I will use a constant D$_{y}$ in most of
the applications in this work. It accounts for the broadening of the
rapidity distributions due to interactions and particle creations, and
it is related to the drift term J(y) by means of a
dissipation-fluctuation theorem which will be used to actually
calculate D$_{y}$.
The drift J(y) determines the shift of the mean rapidities
towards the central value, and I shall discuss linear and nonlinear
forms of this drift function.

In derivations of generalized FPE's from the Boltzmann equation,
a nonlinear equation ($\mu,\nu\ne$1; nonlinear drift function)
could in principle be traced back to the
nonlinearities in the transition probabilities between
single-particle states \cite{wol82}. However, this would not yet include
non-Markovian effects. Instead, (\ref{fpe}) offers the possibility
to describe strong-coupling systems that are beyond the realm of the
Boltzmann equation.

Since the norm of the
rapidity distribution has to be conserved, $\mu$ = 1 is
required here. It is convenient to introduce a
nonextensivity parameter that governs the shape of the
power-law equilibrium distribution, q = 2 - $\nu$ \cite{tsa88}.
In statistical equilibrium,
transverse mass spectra and transverse momentum
fluctuations in relativistic systems at SPS-energies
$\sqrt{s_{NN}}$=17.3 GeV require values
of q very slightly above one, typically
q = 1.038 for produced pions in Pb-Pb \cite{alb00}.
For $q \rightarrow1$, the equilibrium distribution
converges to the exponential Boltzmann form,
whereas for larger values of q (with $q < 1.5$) significantly
broader equilibrium distributions are obtained, and
the time evolution towards them becomes superdiffusive \cite{tsa88,wil00}.

To study rapidity distributions in multiparticle systems at
RHIC energies in a nonequilibrium-statistical framework
\cite{wol99,wols99,lav02,alb02,ryb02,wol03},
I start with q = $\nu$ = 1 corresponding to the standard FPE.
For a linear drift function
 \begin{equation}
J(y)=(y_{eq}- y)/\tau_{y}
\label{dri}
\end{equation}
with the rapidity relaxation time $\tau_{y}$, this is
the so-called Uhlenbeck-Ornstein process \cite{jac96}, applied to the
relativistic invariant rapidity. The equilibrium value is $y_{eq}=0$
in the center-of-mass for symmetric systems, whereas
$y_{eq}$ is calculated from the given masses and
momenta for asymmetric systems. Using $\delta-$function
initial conditions at the beam rapidities $\pm y_{b}$ ($\pm $5.36
at p=100 GeV/c per nucleon), the equation has analytical Gaussian
solutions. The mean values shift in time towards
the equilibrium value according to
\begin{equation}
<y_{1,2}(t)>=y_{eq}[1-exp(-2t/\tau_{y})] \pm y_{b}\exp{(-t/\tau_{y})}.
\label{mean}
\end{equation}
For a constant diffusion coefficient $D_{y}$, the variances of both
distributions have the well-known simple form
\begin{equation}
\sigma_{1,2}^{2}(t)=D_{y}\tau_{y}[1-\exp(-2t/\tau_{y})],
\label{var}
\end{equation}
whereas for a time dependent diffusion coefficient $D_{y}(t)$
that accounts for collective (multiparticle) and memory effects
the analytical expression for the variances becomes more involved \cite{wol02}.
At short times  $t/\tau_{y}<<1$, a statistical description
is of limited validity due to the small number of interactions.
A kinematical cutoff prevents
the diffusion into the unphysical region $|y|>y_{b}$. For larger
values of $t/\tau_{y}$, the system comes closer to statistical
equilibrium such that the FPE is valid.

Since the equation is linear, a superposition of the distribution
functions \cite{wol99} emerging from $R_{1,2}(y,t=0)=\delta(y\mp y_{b})$
\begin{equation}
R(y,t)=\frac{1}{2\sqrt(2\pi\sigma_{1}^{2}(t))}
exp\Bigl[-\frac{(y-<y_{1}(t)>)^{2}}{2\sigma_{1}^{2}}\Bigr]+
\frac{1}{2\sqrt(2\pi\sigma_{2}^{2}(t))}
exp\Bigl[-\frac{(y-<y_{2}(t)>)^{2}}{2\sigma_{2}^{2}}\Bigr]
\label{fpesol}
\end{equation}
yields the exact solution (normalized to 1).
With the the total number of net baryons (or protons, depending on the
experimental results)
N$_{1}$+N$_{2}$ the rapidity density distribution of net baryons
(protons) becomes
\begin{equation}
\frac{dN(y,t=\tau_{int})}{dy}=N_{1}R_{1}(y,\tau_{int})+N_{2}R_{2}(y,\tau_{int}).
\label{norm}
\end{equation}
The value of $t/\tau_{y}$ at the interaction time t=$\tau_{int}$
(the final time in the integration of (\ref{fpe})) is determined by
the peak positions \cite{wol99} of the experimental distributions.
The same approach has also been applied successfully
by Biyajima et al. to produced
particles at RHIC energies \cite{biy02}, although the initial conditions
are less straightforward for produced particles, as compared to
participant baryons (they use $\delta$-function initial conditions
also for produced hadrons).

The microscopic physics is
contained in the diffusion coefficient. Macroscopically,
the transport coefficients are related
to each other in the weak-coupling limit (D$_{y}^{w}$) through the
dissipation-fluctuation theorem (Einstein relation)
with the equilibrium temperature T
\begin{equation}
D_{y}^{w}=\alpha\cdot T\simeq f(\tau_{y},T).
\label{dft}
\end{equation}
In \cite{wol99} I have obtained the analytical result for $D_{y}^{w}$
as function of T and $\tau_{y}$ from the condition
that the stationary solution of (\ref{fpe}) is equated
with a Gaussian approximation to the thermal equilibrium
distribution in y-space (which is not exactly Gaussian, but
very close to it) as
\begin{equation}
D_{y}^{w}(\tau_{y},T)=\frac{1}{2\pi\tau_{y}}\Bigl[C(\sqrt{s},T)\cdot
(1+2\frac{T}{m}
+2(\frac{T}{m})^2)\Bigr]^{-2}exp(\frac{2m}{T})
\label{diffwc}
\end{equation}
with $C(\sqrt{s},T)$ given in \cite{wols99} in closed form:
For a given equilibrium distribution, the rapidity diffusion
coefficient is then determined by the dissipative constant $\tau_{y}$.
The proton mass is m, and $C(\sqrt{s},T)$ ensures
that the corresponding thermal distribution function
$R_{th}$ is normalized to 1 for each temperature T
\begin{equation}
\int    \limits_{-\infty}^{+\infty}R_{th}(y)dy = 1.
\label{lim}
\end{equation}
 This yields
\begin{equation}
C(\sqrt{s},T)=y_{b}\Bigl[ \int
\limits_{-\infty}^{+\infty}\Bigl(1+2\chi_{T}(y)+
2\chi_{T}(y)^{2}\Bigr)\exp\left(\frac{-1}{\chi_{T}(y)}\right)dy\Bigr]^{-1}
\label{cpt}
\end{equation}
with the beam rapidity $y_{b}$ in the center-of-mass, and
\begin{equation}
\chi_{T}(y)=\frac{T}{m\cosh(y)}.
\label{chi}
\end{equation}
(Note that in \cite{wols99} the result is written for the beam
rapidity in the laboratory system y$_{1}$=5.83 at
the SPS-energy for Pb-Pb, corresponding here to y$_{b}=\pm$2.195
in the center-of-mass).

At fixed beam rapidity $y_{b}$, the diffusion
coefficient displays the expected
behavior, namely, it rises with increasing temperature of the
corresponding equilibrium distribution as in the
Einstein relation; the rise is almost linear with T.
This allows
to maintain the linearity
of the model and hence, to solve the FPE analytically,
although small corrections are to be expected. They cause
minor deviations in the calculated rapidity distributions that
are within the size of the
error bars of the experimental data at SPS energies.
\section{Strong-coupling diffusion: Memory effects}
In the linear model, net baryon rapidity spectra at
low SIS-energies (about 1 GeV per particle) are well reproduced,
whereas at AGS and SPS energies I find discrepancies
\cite{wols99} to the
data that rise strongly with $\sqrt{s}$. The origin are most likely
strong-coupling effects at high energy: the time between two subsequent
interactions becomes smaller than the duration of an individual
interaction, such that the system becomes non-Markovian
and develops memory-effects. In a schematic approach that
serves to outline the effect analytically, one may account for the
system memory by
a time-dependent diffusion coefficient
through a relaxation ansatz
\begin{equation}
\frac{\partial D_{y}(t)}{\partial t}=-\frac{1}{\tau_{s}}
\Bigl[D_{y}(t)-D_{y}^{w}\Bigr].
\label{dit}
\end{equation}
Here the diffusion
coefficient for weak coupling $D_{y}^{w}$ is
approached for times $t\gg\tau_{s}$ larger
than the time $\tau_{s}$ that is characteristic
for strong coupling - when all secondary particles
have been created. $D_{y}^{w}$ is well defined in
terms of the temperature of the corresponding
 equilibrium distribution, and the particle mass
 as discussed above. However, for short times
 in the initial phase of the collision before and
 during particle production, the strong-coupling
 diffusion coefficient $D_{y}^{s}$ dominates
 and enhances the diffusion in y-space beyond
 the weak-coupling value. This enhancement
 rises strongly with incident energy as shown
 in \cite{wols99}. It is decisive for a proper representation
 of the available data for relativistic heavy-ion
 collisions at and beyond SPS energies. The relaxation equation
 for the rapidity
 diffusion coefficient with the strong-coupling
 value D$_{y}^{s}$ as an initial condition can be solved as
\begin{equation}
D_{y}(t)=D_{y}^{w}[1-exp(-\frac{t}{\tau_{s}})]+
D_{y}^{s}exp(-\frac{t}{\tau_{s}}).
\label{diffsc}
\end{equation}
A lower limit for the characteristic time for
 strong coupling $\tau_{s}$ can be estimated
 from the time delay of particle production
 due to quantum coherence \cite{gal99}
which yields about 0.4 fm/c at SPS energies,
 or simply from the uncertainty principle,
which gives about the same value for a pion.
The strong-coupling time $\tau_{s}$ is,
however, much larger since it refers to
 many particles that are created from
the available relativistic energy. The
 transverse energy relaxation time \cite{wol96} -
which is of the order of, but slightly smaller
than the interaction time at SPS energies -
gives a reasonable estimate for the strong-coupling
time such that
\begin{equation}
\tau_{s}<\tau_{int}<\tau_{y}
\end{equation}
at SPS-energies. Not much is presently known about the
 strong-coupling diffusion coefficient $D_{y}^{s}$ except
 that it is significantly larger than its weak-coupling
counterpart at these energies. Unlike $D_{y}^{w}$, it can
 not be derived from the known equilibrium distribution.
 Moreover, energy and momentum-conservation are not
fulfilled in the strong-coupling phase of the collision.
Leaving a microscopic model to the future, I presently adjust in the
linear model
D$_{y}(t=\tau_{int})$ to the data as outlined in the next section.
Due to the
 strong coupling at short times t/$\tau_{y}\ll 1$,
D$_{y}$
 is initially far above the weak-coupling value D$_{y}^{w}$
 (\ref{diffwc}), and
 it remains there throughout the interaction time at SPS energies,
 even though it drops cslightly until freezeout occurs.
The corresponding variance in rapidity space is obtained analytically
through a solution of the corresponding differential equation
for the variances with the
time-dependent diffusion coefficient (\ref{diffsc}) contained in
the inhomogeneity
\begin{equation}
\frac{d}{dt}\sigma_{y}^{2}(t)+2\sigma_{y}^{2}(t)=2D_{y}(t)\tau_{y}.
\end{equation}
It can be written as a sum of weak- and
strong-coupling contributions, respectively
\begin{equation}
\sigma_{y}^{2}(t)=\sigma_{w}^{2}(t)+
\sigma_{s}^{2}(t)
\end{equation}
with
\begin{equation}
\sigma_{w}^{2}(t)=D_{y}^{w}\tau_{y}exp(-2t/\tau_{y})
\Bigl\{\Bigl[exp(2t/\tau_{y})-1\Bigr]+\frac{2\tau_{s}}{\tau_{y}
-2\tau_{s}}\Bigl[exp(-t/\tau_{y}\frac{\tau_{y}-2\tau_{s}}
{\tau_{s}})-1\Bigr]\Bigr\}
\end{equation}
and
\begin{equation}
\sigma_{s}^{2}(t)=D_{y}^{s}\tau_{y}exp(-2t/\tau_{y})
\Bigl[\frac{-2\tau_{s}}{\tau_{y}-2\tau_{s}}\Bigr]
\Bigl[exp(-t/\tau_{y}\frac{\tau_{y}-2\tau_{s}}{\tau_{s}})
-1\Bigr].
\end{equation}\\
In the limit of $\tau_{s}\rightarrow0$ the fluctuations
 due to strong coupling vanish, and the remaining weak-coupling
 result for the variance of the rapidity distribution function
 attains the familiar form (\ref{var})
\begin{equation}
\sigma_{y}^{2}(t)\rightarrow D_{y}^{w}\tau_{y}
\Bigl[1-exp{(-2t/\tau_{y})}\Bigr] {\quad}for {\quad} \tau_{s}\rightarrow0.
\end{equation}\\
Results for the standard deviations in Y-space (Y=y/y$_{b}$) of the superposed
 FPE-solutions (\ref{fpesol}) that build up the baryon rapidity
distributions in the relativistic system 158 GeV/c
 per nucleon Pb-Pb are shown in Fig.2. The
strong-coupling value (upper curve) would persist
 if no particles were created. Due to particle creation,
 the actual fluctuations pass a maximum and gradually
 approach the weak-coupling result for large times.
 The interaction time (here for a central collision)
 corresponds approximately to the maximum: freezeout
 occurs for large fluctuations in rapidity at SPS energies
 and beyond.
\section{From SPS to RHIC energies}
At SPS energies \cite{ma95}, the discrepancy between
weak-coupling result and data has recently been confirmed independently
in a numerical calculation \cite{lav02,alb02} based on a nonlinear drift
\begin{equation}
J(y)=-\alpha\cdot m_{\perp}sinh(y)\equiv -\alpha\cdot p_{\parallel}
\label{drinl}
\end{equation}
with the transverse mass $m_{\perp}=\sqrt{m^2 + p_{\perp}^2}$,
and the longitudinal momentum $p_{\parallel}$.
Together with the dissipation-fluctuation theorem (\ref{dft}), this
yields exactly the Boltzmann distribution as the stationary
solution of (\ref{fpe}) for $\nu=q=1$. The corresponding numerical
solution with $\delta-$function initial conditions
at the beam rapidities is, however, only approximately correct
since the superposition principle is not strictly valid
for a nonlinear drift. Still, the numerical result shows
almost the same large discrepancy
between data and theoretical rapidity distribution
as the linear model. In a q=1 framework, the net
proton distribution in Pb-Pb at the highest SPS energy requires
a rapidity width coefficient $\sqrt{D_{y}\tau_{y}}$
that is enhanced beyond the
theoretical value (\ref{dft}) by a factor of
$g(\sqrt{s})\simeq2.6$ due to memory and collective effects
\cite{wol99,wols99,wol02}, Fig.1.

Alternatively, a
transition to nonextensive statistics \cite{tsa88,alb00,wil00}
maintaining the weak-coupling diffusion coefficient
from (\ref{dft}) requires a value of q that
is significantly larger than one. In an approximate
numerical solution of (\ref{fpe}) with the nonlinear drift
(\ref{drinl}),
q=1.25 has been determined for the net-proton
rapidity distribution in Pb-Pb collisions at the SPS \cite{lav02,alb02}.
The only free parameter is q, whereas in the
linear q=1 case the strong-coupling enhancement
of $D_{y}$ beyond (\ref{dft})  is
the only parameter.

This value of q in the nonlinear model
is considerably larger than the result
$q(\sqrt{s_{NN}})$=1.12 extrapolated from Wilk et
al. \cite{ryb02} at the SPS-energy
$\sqrt{s_{NN}}$
= 17.3 GeV. Here, the relativistic diffusion
approach is applied
to produced particles in proton-antiproton collisions
in the energy range $\sqrt{s}$ = 53 GeV - 1800 GeV,
and used to predict LHC-results. The nonlinearity $q>1$
appears to be an essential feature of the
 $p\bar p$ data. The larger value
of q in heavy systems as compared to $p\bar p$
at the same NN-center-of-mass energy
emphasizes the increasing superdiffusive
effect of multiparticle
collisions both between participants,
and between participants
and produced particles. It is, however, conceivable
that both a violation of (\ref{dft}) due to memory effects, and $q>1$
have to be considered in a complete description.

The Au+Au system at RHIC energies is then investigated first in
the linear q=1 model for  central collisions (10 per cent
of the cross-section).  Based on the experience at AGS
and SPS energies \cite{wol99,wols99,wol02},
it is expected that the nonequilibrium net-proton rapidity
spectrum calculated with
the weak-coupling dissipation-fluctuation theorem (\ref{dft})
underpredicts the widths of the nonequilibrium
fractions of the experimental distribution significantly.
At RHIC energies, the precise value of the enhancement due
to multiparticle effects remains somewhat uncertain at present since the
largest-rapidity experimental points are on the edge of the nonequilibrium
distribution \cite{lee02,bea02}. Typical solutions of the linear FPE
for various values of t/${\tau_{y}}$ representing the diffusive time evolution
of the baryonic system due to interactions and
particle creations are shown in Fig.3.

In the comparison with the BRAHMS data \cite{lee02} shown in
Fig.4 (bottom), the temperature T=170 MeV is taken from a thermal fit
of charged antiparticle-to-particle ratios
in the Au+Au system at 200 GeV per nucleon \cite{bec01,bea02a},
and the theoretical value of the rapidity width coefficient
calculated from the analytical expression (6) is
$\sqrt{D_{y}\tau_{y}}=7.6\cdot 10^{-2}.$ The weak-coupling
nonequilibrium distributions without enhancement due
to multiparticle effects (dotted curves in Fig.4)
are by far too narrow to represent the
data, justifying the term "anomalous" for the
experimental result.

The calculated distributions become even slightly narrower
when T is lowered in order to account for the fact that
the equilibrium temperature in the diffusion model should
be associated with the kinetic freeze-out temperature
(which is not yet precisely known at RHIC), rather
than the chemical freeze-out temperature. As has been
discussed in \cite{wols99}, the weak-coupling rapidity diffusion
coefficient is proportional to the temperature as in the
theory of Brownian motion \cite{ein05}, $D_{y}\propto T/{\tau_{y}}$.
Hence, lowering the temperature by 40 MeV reduces the widths
of the nonequilibrium distributions by
12 per cent - which is hardly visible on the scale of Fig.4.

It was shown in \cite{lav02} for SPS results that the discrepancy
between nonequilibrium weak-coupling result and data persists in case of the
nonlinear drift (\ref{drinl}) that yields the exact Boltzmann-Gibbs
equilibrium solution for q=1
\begin{equation}
E\frac{d^{3}N}{d^{3}p}=\frac{d^{3}N}{dy\cdot
m_{\perp}dm_{\perp}d\phi}\propto E\cdot exp(-E/T).
\label{bge}
\end{equation}
Hence, it is expected that the nonlinear drift (\ref{drinl}) does not improve
the situation in the q=1 case at RHIC energies.

Instead, an enhancement of the weak-coupling
rapidity width coefficient
by a factor of $g(\sqrt{s})\simeq3.7$
due to collective and
memory effects in the system
corresponding to a violation of (\ref{dft}) yields
a good reproduction of the nonequilibrium
contributions with
$\tau_{int}/\tau_{y}$=0.26 .
However, the
midrapidity valley that is present in the data is completely
absent in the extensive nonequilibrium q=1 case, solid curves
in Fig.4 (bottom).

This remains true in the
nonextensive case ($1<q<1.5$), with an approximate distribution
function \cite{tsa88,wil00,lav02,alb02,ryb02,wol03} that is given by a linear
superposition of power-law
solutions of (\ref{fpe})
\begin{equation}
R_{1,2}(y,t)=[1-(1-q)\frac{m_{\perp}}{T}
cosh(y-<y_{1,2}(t)>)]^{\frac{1}{1-q}}.
\label{nesol}
\end{equation}
The dashed curves in Fig.4 (bottom) show the result for q=1.4,
T=170 MeV and a mean
transverse mass $<m_{\perp}>$=1.2 GeV.
This solution is far from the nonextensive
equilibrium distribution which would be reached for
$<y_{1,2}(t\rightarrow\infty)>=y_{eq}$, and it
is significantly below the midrapidity data. The
result is even worse for larger values of $m_{\perp}$.
In contrast, the Pb-Pb data at SPS energies \cite{app99} are well
described both in the linear model \cite{wol99} (Fig.6, top) and in the
nonlinear case (cf. \cite{lav02} for results with a time-dependent
temperature and an integration over transverse mass).
\section{Local deconfinement at RHIC}
It turns out, however, that the RHIC data can be
interpreted rather precisely in the linear q=1 framework
with the conjecture that a fraction of
$Z_{eq}\simeq 22$ net protons near midrapidity
reaches local statistical equilibrium
in the longitudinal degrees of freedom, Figs.4(top),5. The variance of
the equilibrium distribution $R_{eq}(y)$ at midrapidity is broadened
as compared to the Boltzmann result
due to collective (multiparticle) effects by
the same factor that enhances the
theoretical weak-coupling diffusion coefficient derived from
(\ref{dft}), shaded areas in Figs.4,5.
This may correspond to a longitudinal expansion (flow)
velocity of the locally equilibrated subsystem
as accounted for in hydrodynamical descriptions.
In the nonextensive model, the corresponding local equilibrium distribution
is broadened (blue-shifted) according to $q\simeq1.4$.
Here the enhanced value of q appears
as a convenient parametrization of collective expansion
("longitudinal flow").

Microscopically, the baryon transport over 4-5
units of rapidity to the
equilibrated midrapidity region is not only due to
hard processes acting on single valence (di)quarks
that are described by perturbative QCD,
since this yields insufficient stopping. Instead,
additional processes such as the nonperturbative
gluon junction mechanism \cite{ros80} are necessary
to produce the observed central valley. This
may lead to substantial stopping even at LHC energies
where the separation of nonequilibrium and equilibrium
net baryon fractions in rapidity space is expected
to be even better than at RHIC. In the late
thermalization stage \cite{ber01},
nonperturbative approaches to QCD thermodynamics
are expected to be important.

Recent work indicates that one may account
for the observed stopping in heavy-ion collisions at SPS and RHIC
energies with string-model
parameters determined from hadron-hadron collisions \cite{cap02}. If this
was confirmed, the corresponding rapidity distributions would not be
considered to be anomalous from a microscopic point of view.
However, this view does not offer a distinction between
nonequilibrium and equilibrium contributions to the net baryon
rapidity spectra, which both exist at RHIC energies, and
are anomalously broadened.

Macroscopically, the complete solution of (\ref{fpe})
in the q=1 case is a linear superposition
of nonequilibrium and local equilibrium distributions (Fig.6, bottom)
\begin{equation}
R(y,t=\tau_{int})=R_{1}(y,\tau_{int})+R_{2}(y,\tau_{int})
+R_{eq}^{loc}(y).
\label{sol}
\end{equation}
with the same enhancement factor $g(\sqrt{s})$ due to multiparticle
(collective) effects for all three distributions. The net-baryon rapidity
distribution becomes
\begin{equation}
\frac{dN(y,t=\tau_{int})}{dy}=N_{1}R_{1}(y,\tau_{int})+N_{2}R_{2}(y,\tau_{int})
+N_{eq}R_{eq}^{loc}(y)
\label{normloc}
\end{equation}
with the number N$_{eq}$ of net baryons (here: net protons) in local equilibrium
near mid-rapidity, and N$_{1}$ + N$_{2}$ + N$_{eq}$ equal to the total
number of net baryons (158 net protons for central Au-Au).
This yields a good representation of the preliminary
BRAHMS data \cite{lee02}.  (In the q$>1$ case, the corresponding
solution is questionable because
the superposition principle is violated).
Based on (\ref{normloc}), the transition from net-proton rapidity
spectra with a central plateau in Pb-Pb at the lower SPS energies \cite{wol02},
via a double-humped
distribution at the maximum SPS energy \cite{app99,wol99,lav02,alb02,wol02}
to the central
valley at RHIC \cite{lee02,wol03} is well understood.
It has not yet been possible
to identify a locally equilibrated subsystem of net baryons
at midrapidity below RHIC-energies, although it cannot
be excluded that it exists.
At SPS energies, the data \cite{app99} are well described by the
nonequilibrium distributions, and it is much more difficult
(and probably impossible) to identify a locally equilibrated
component because the relevant rapidity region is comparatively small,
and an equilibrated contribution cannot be separated from the
nonequilibrium components in rapidity space.
In  $p\bar p$-collisions at
$\sqrt{s}=$53-900 GeV,
no convincing signatures of a phase transition were found \cite{gei89}.

Most remarkably, Figs.4-6 suggest that in central Au+Au collisions
at $\sqrt{s_{NN}}$= 200 GeV
there is no continuous transition from the nonequilibrium
to the equilibrium contribution in net-proton rapidity spectra
as function of time.
The central valley in net-proton rapidity
spectra at RHIC energies could thus
be used as an indicator for partonic processes
that lead to a baryon transfer over more than 4
units of rapidity, and for quark-gluon plasma formation.
The discontinuity may well be due to a sudden enhancement in the number
of degrees of freedom as encountered in the
deconfinement of participant partons, which enforces a very rapid
local equilibration in a fraction of the system as indicated in
Figs.4,5
through a discontinuity in the time evolution of the solutions of the FPE,
which are afterwards very close to the equilibrium result.
Here, the sudden enhancement in the number of degrees of freedom by a factor
of about 6 in the quark-gluon plasma as compared to the hadronic phase
\cite{djs03} is modelled by a corresponding increase in time, Fig.5.

The linear RDM with three sources \cite{wol03} including a local
equilibrium fraction near midrapidity has recently also been adopted
by Biyajima et al. \cite{biya03} for produced particles in analyses
of BRAHMS \cite{bear01} and PHOBOS \cite{bac01} pseudorapidity data at
$\sqrt{s}=$130 GeV and 200 GeV, using the proper Jacobian to
transform from y to $\eta$=-lntan($\theta/2$). In particular, the
PHOBOS data \cite{bac01} at 130 (200) GeV and 0-6 per cent centrality are
well reproduced with 3134 (3858) charged hadrons in the central source
indicative of an equilibrated qgp, whereas only 896 (1102) charged
hadrons reside in the non-equilibrated fragmentation regions.

These very recent results for produced charged hadrons are, however, still
somewhat ambiguous in view of (1) the uncertainties regarding the
initial conditions for produced particles in rapidity space, and (2)
the difficulties to clearly separate fragmentation and central regions
for produced particles. Nevertheless, the RDM-analyses of produced
hadrons by Biyajima et al. \cite{biya03} provide additional evidence
for local equilibrium near midrapidity.
\section{Conclusion}
To conclude, I have interpreted recent results for
central Au-Au collisions at RHIC energies in
a Relativistic Diffusion Model (RDM) for
multiparticle interactions based on the interplay of
nonequilibrium and local equilibrium ("thermal") solutions.
In the linear version of the model, analytical results for the rapidity
distribution of net protons in central collisions have been
obtained. The anomalous enhancement of the diffusion in rapidity
space as compared to the expectation from the
weak-coupling dissipation-fluctuation theorem due to
high-energy multiparticle effects has been discussed
using extensive and nonextensive statistics.

A significant
fraction of about 14 per cent of the net protons reaches
local statistical equilibrium
in a fast and discontinuous transition which is likely to
indicate parton deconfinement. The precise amount of protons
in local equilibrium is related to the experimental
value of the rapidity density close to y=0 and hence, possible changes in the
final data will affect the percentage. It has not yet been
possible to isolate a corresponding fraction of longitudinally
equilibrated net protons in the Pb-Pb system
at SPS energies. Since no signatures of a transition to the
quark-gluon plasma have been observed in $p\bar p$-collisions,
Quark-Matter formation is clearly
a genuine many-body effect occuring
only in heavy systems at sufficiently high energy density.
Consequently, a detailed investigation of
the flat midrapidity valley found at RHIC, and of its energy
dependence is very promising.

\newpage
\Large\bf
Figure captions
\normalsize\rm
\begin{description}
\item[FIG. 1:]
Net proton rapidity spectra in the Relativistic Diffusion Model
(RDM), solid curves, at the lower SPS momentum of 40 GeV/c
per particle (top; data not yet available), at 158 GeV/c
(compared to NA 49 data \cite{app99}; cf. Fig.6
for error bars) and at the highest RHIC
energy $\sqrt{s_{NN}}$ = 200 GeV (compared to
preliminary BRAHMS data \cite{lee02}).
The variable Y=y/y$_{b}$ renormalizes the y-distributions
in the center-of-mass
with the beam rapidities $\pm$y$_{b}$.
Dashed lines are thermal equilibrium
results without expansion. The transition from bell-shaped to
double-humped is clearly shown in the RDM.
\item[Fig. 2.]
Standard deviations $\sigma_{Y}(t)$ of the solutions of the Fokker-Planck
equation that build up the Pb-Pb baryon rapidity distribution
at SPS momenta of 158 GeV/c per particle in Y-space (Fig.1). The
upper curve is the strong-coupling result. Due to particle
creation, the actual fluctuations (middle curve) gradually
approach the weak-coupling result, lower curve. The interaction
 time $\tau_{int}/\tau_{y}$ (arrow)
approximately corresponds to the maximum of the
fluctuations in rapidity space.
\item[FIG. 3:]
Analytical solutions of the Relativistic Diffusion Model for various
values of t/${\tau_{y}}$ representing the diffusive time evolution
of the baryonic system due to interactions and
particle creations for central Au-Au at RHIC energies.
Here the net proton content is 136 protons. This subsystem remains in
a non-equilibrium state in the actual experiment, cf. Fig.6.
\item[FIG. 4:]
Nonequilibrium contributions to the net-proton rapidity
spectra of Au-Au at $\sqrt{s_{NN}}$ = 200 GeV in
the Relativistic Diffusion Model (RDM) with an equilibrium
temperature of T=170 MeV (bottom). The solid curve is obtained
in the linear model (q=1). Its width is ("anomalously")
enhanced as compared to
the theoretical weak-coupling value (dotted curves) by
$g(\sqrt{s})=3.7$ due to multiparticle effects \cite{wols99,wol02}
according to the preliminary BRAHMS data \cite{lee02}, squares.
The net proton content is 158.
The dashed curve corresponds to q=1.4 and $<m_{\perp}>$=1.2 GeV
in the nonlinear model. Reasonable agreement with
the midrapidity data requires that 14 per cent
of the net protons reach local thermal equilibrium in a
discontinuous transition, top (dashed curve: thermal equilibrium
result for 22 protons; shaded area: broadening due to multiparticle
effects with $g(\sqrt{s})=3.7$).
\item[FIG. 5:]
Analytical solutions of the Relativistic Diffusion Model for various
values of t/${\tau_{y}}$ representing the time evolution
of a local subsystem of the net baryons in central Au-Au at RHIC
energies with a sudden enhancement
of the number of degrees of freedom due to deconfinement,
and the subsequent fast local equilibration. The net proton content
of this local subsystem is 22. The shaded area is the local
equilibrium distribution centered at midrapidity.
\item[FIG. 6:]
Net-proton rapidity spectra for central collisions of
Au-Au at $\sqrt{s_{NN}}$ = 200 GeV consist
of two nonequilibrium components (solid peaks, bottom; 136 protons)
plus a local equilibrium contribution at T=170 MeV.
The shaded area shows its broadening due to collective
(multiparticle) effects by the same factor $g(\sqrt{s})=3.7$
as the nonequilibrium fractions.
After hadronization, it contains
$Z_{eq}\simeq22$ protons. Superposition creates the
flat valley near midrapidity (bottom) in agreement with the
preliminary BRAHMS data points \cite{lee02}.
Diamonds include $\Lambda$ feed-down corrections at y=0 (17.5 per cent)
and y=2.9 (20 per cent), respectively.
Arrows indicate the beam rapidities $\pm y_{b}$.
At SPS energies (top), NA49 data \cite{app99}
for central Pb-Pb events (5 per cent) including $\Lambda$ feed-down corrections
are compared with the pure nonequilibrium
result of the linear model \cite{wol99,wol02}. Here, the net
proton content is 164.
\end{description}
\newpage
\vspace{1cm}
\includegraphics[bb=10 80 440 670]{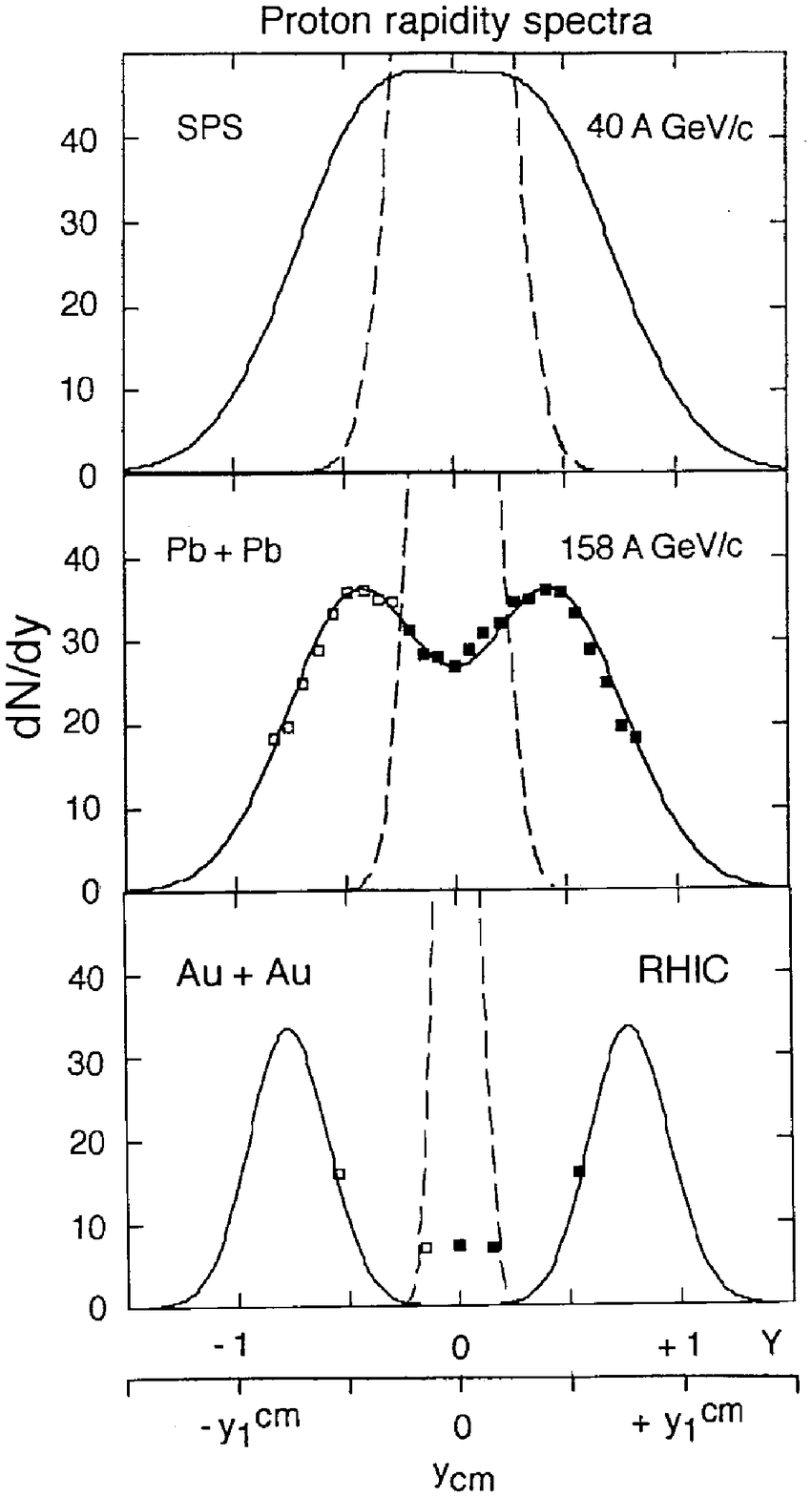}
\newpage
\vspace{1cm}
\includegraphics[bb=10 80 440 670]{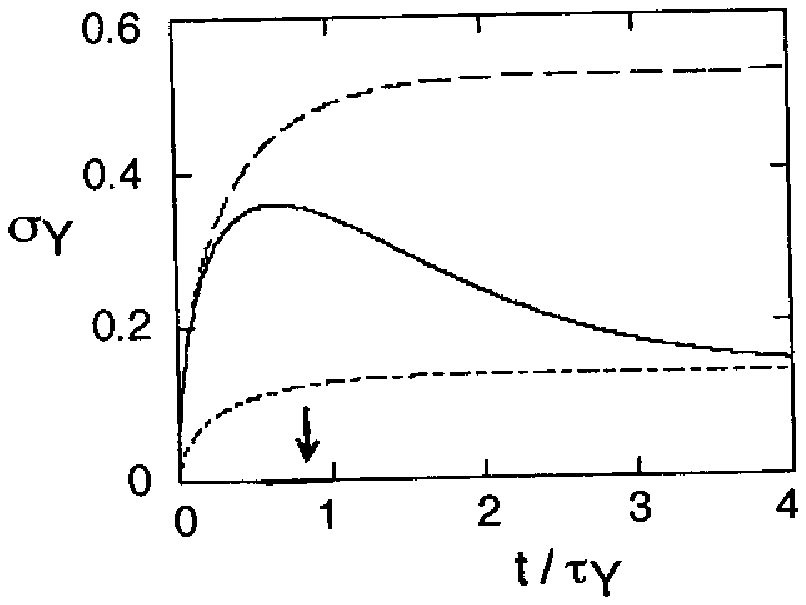}
\newpage
\vspace{1cm}
\includegraphics[width=14cm]{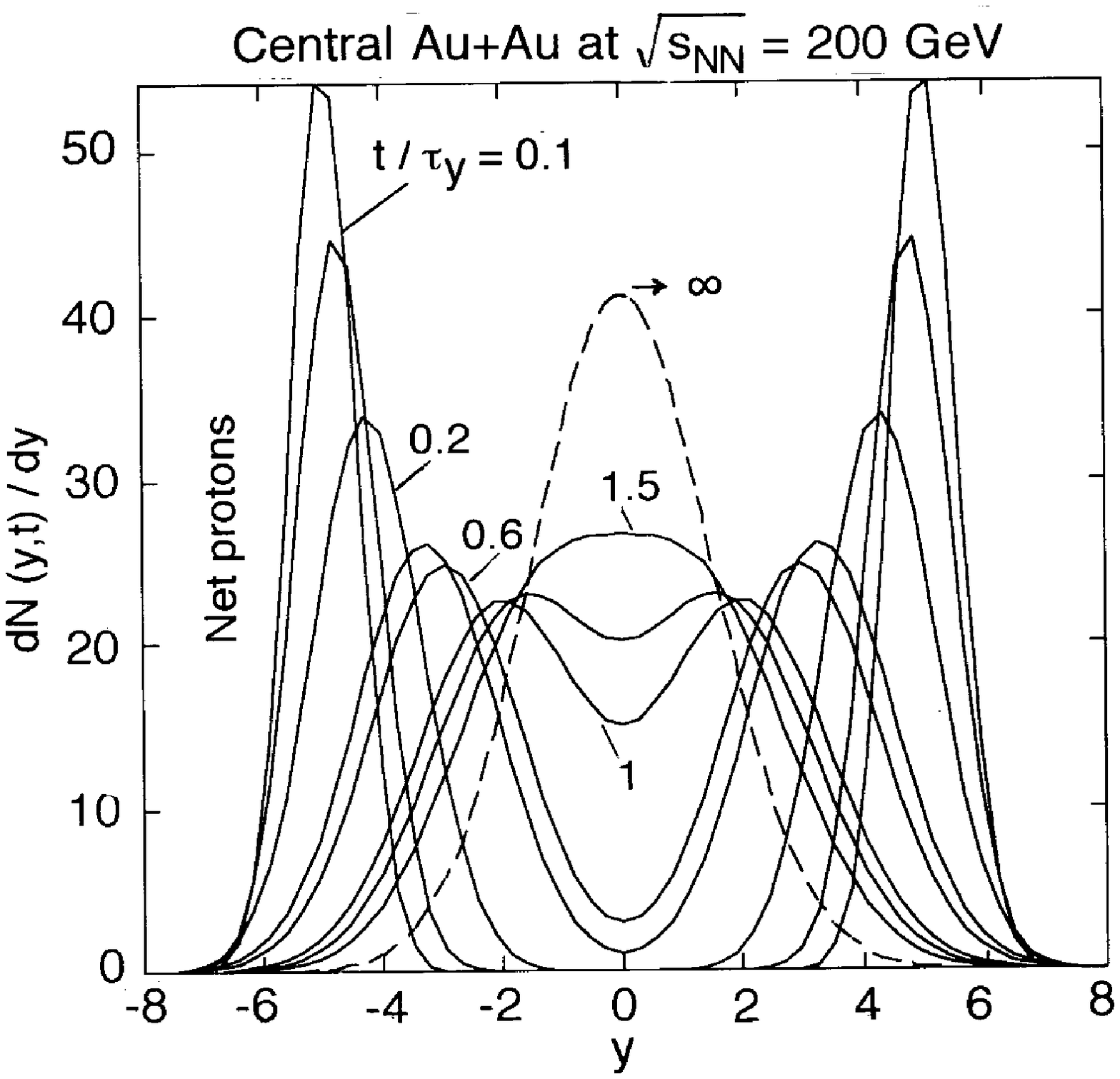}
\newpage
\vspace{1cm}
\includegraphics[height=14cm]{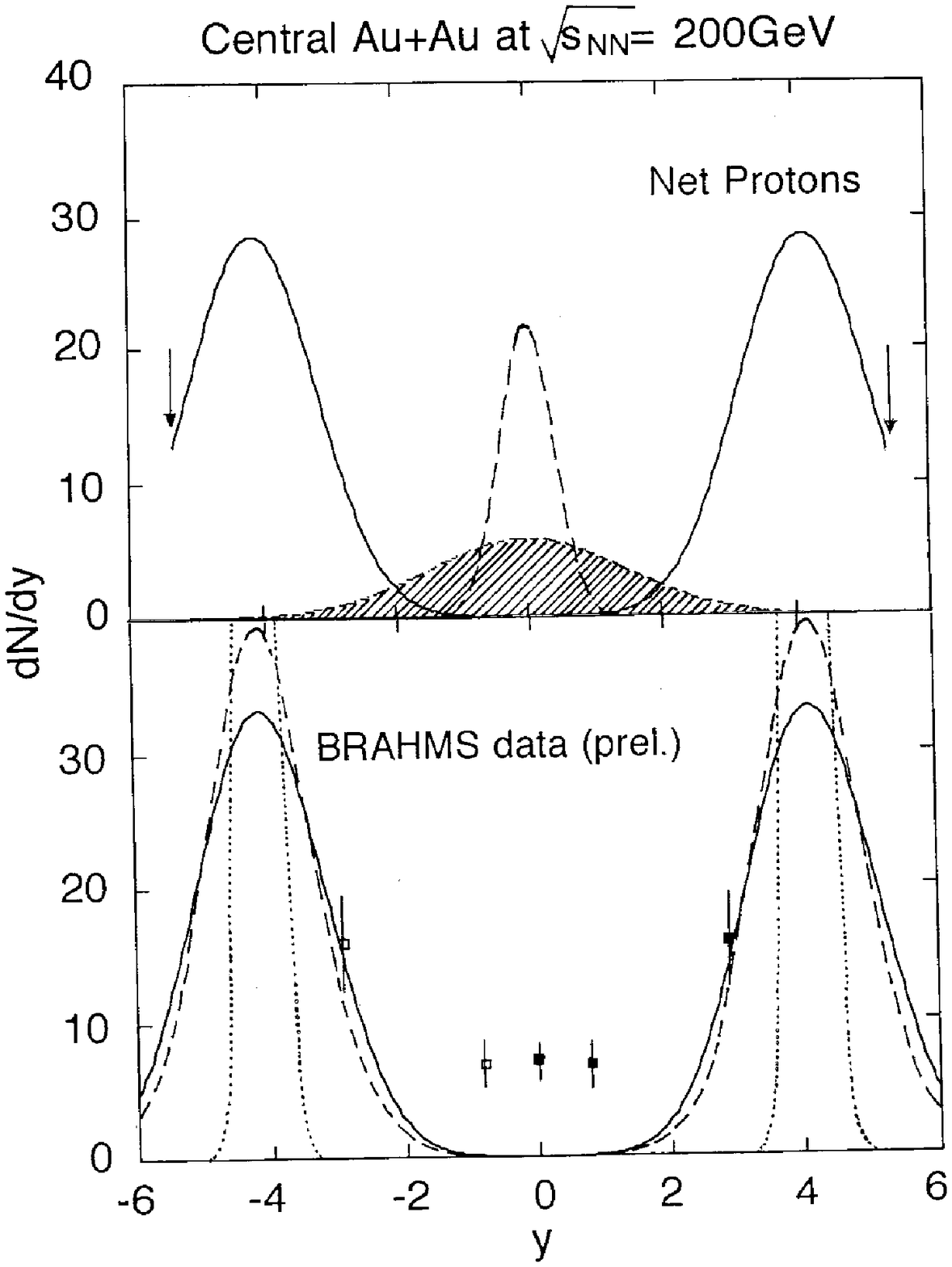}
\newpage
\vspace{1cm}
\includegraphics[width=14cm]{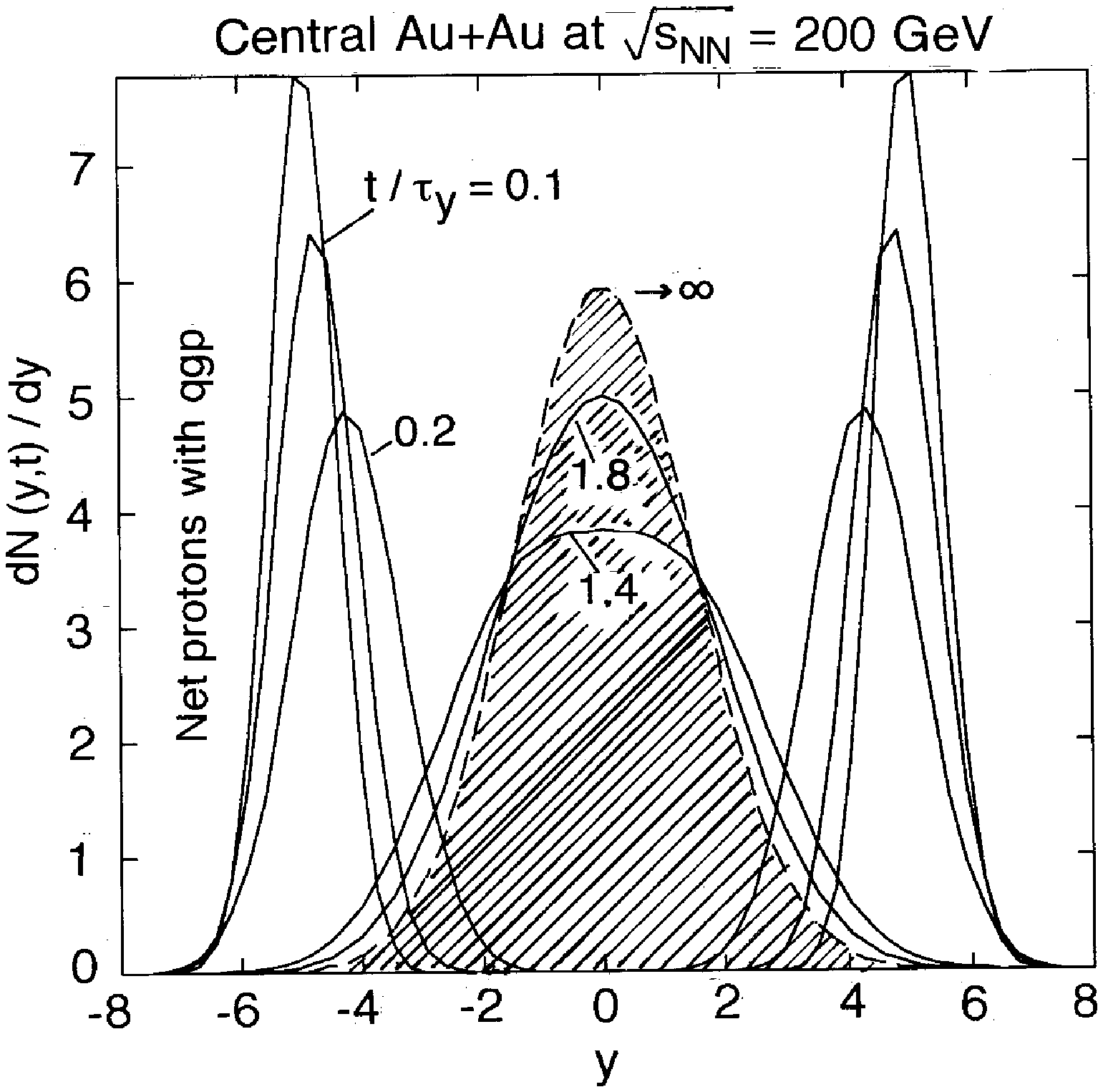}
\newpage
\vspace{1cm}
\includegraphics[height=14cm]{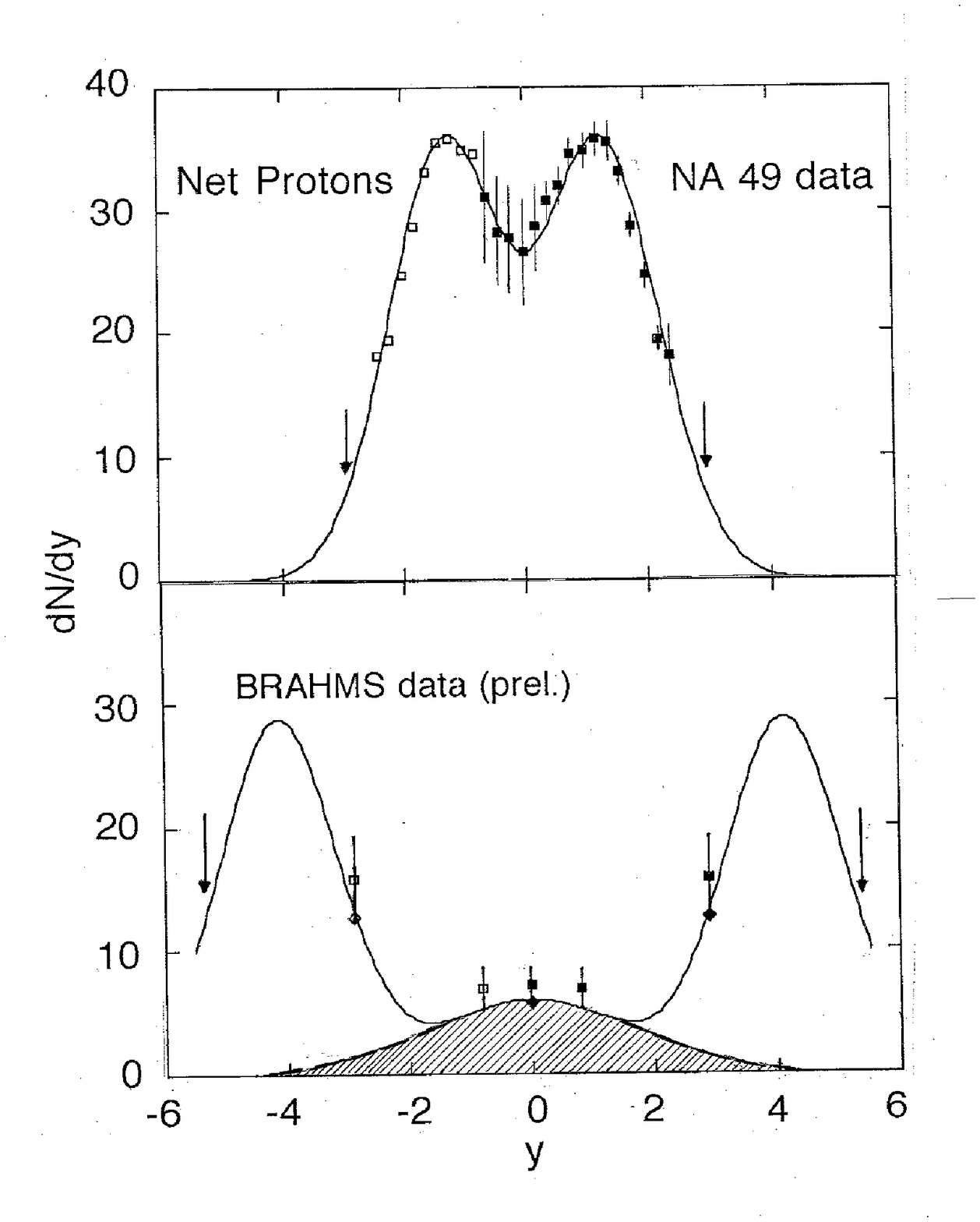}
\end{document}